\documentclass[english,prl,priprint,twocolumn,superscriptaddress]{revtex4-1}
\usepackage[LGR,T1]{fontenc}
\usepackage[latin9]{inputenc}
\usepackage{xcolor}
\usepackage{pdfcolmk}
\usepackage{booktabs}
\usepackage{mathrsfs}
\usepackage{amsmath}
\usepackage{amssymb}
\usepackage{graphicx}
\PassOptionsToPackage{normalem}{ulem}
\usepackage{ulem}

\makeatletter

\DeclareRobustCommand{\greektext}{%
  \fontencoding{LGR}\selectfont\def\encodingdefault{LGR}}
\DeclareRobustCommand{\textgreek}[1]{\leavevmode{\greektext #1}}
\ProvideTextCommand{\~}{LGR}[1]{\char126#1}

\newcommand{\lyxmathsym}[1]{\ifmmode\begingroup\def\b@ld{bold}
  \text{\ifx\math@version\b@ld\bfseries\fi#1}\endgroup\else#1\fi}

\providecommand{\tabularnewline}{\\}
\providecolor{lyxadded}{rgb}{0,0,1}
\providecolor{lyxdeleted}{rgb}{1,0,0}

\DeclareRobustCommand{\lyxsout}[1]{\ifx\\#1\else\sout{#1}\fi}

\makeatother

\usepackage{babel}
\begin{document}
\title{Chiral Majorana Hinge Modes in Superconducting Dirac Materials}
\author{Bo Fu}
\affiliation{Department of Physics, The University of Hong Kong, Pokfulam Road, Hong Kong, China}
\author{Zi-Ang Hu}
\affiliation{Department of Physics, The University of Hong Kong, Pokfulam Road, Hong Kong, China}
\author{Chang-An Li}
\affiliation{Institute for Theoretical Physics and Astrophysics, University of Wurzburg, D-97074
Wurzburg, Germany}
\author{Jian Li}
\affiliation{School of Science, Westlake University, 18 Shilongshan Road, Hangzhou 310024, Zhejiang
Province, China}
\affiliation{Institute of Natural Sciences, Westlake Institute for Advanced Study, 18 Shilongshan
Road, Hangzhou 310024, Zhejiang Province, China}
\author{Shun-Qing Shen}
\email{sshen@hku.hk}

\affiliation{Department of Physics, The University of Hong Kong, Pokfulam Road, Hong Kong, China}
\date{\today}
\begin{abstract}
Chiral Majorana hinge modes are characteristic of a second-order topological superconductor
in three dimensions. Here we systematically study pairing symmetry in the point group
$D_{2h}$, and find that the leading pairing channels can be of $s$-, $d$-, and
$s+id$-wave pairing in Dirac materials. Except for the odd-parity $s$-wave pairing
superconductivity, the $s+id$-wave pairing superconductor is topologically nontrivial
and possesses Majorana hinge and surface modes. The chiral Majorana hinge modes can
be characterized by a winding number of the quadrupole moment, or quantized quadruple
moment at the symmetrically invariant point. Our findings suggest the strong spin-orbital
coupling, crystalline symmetries and electron-electron interaction in the Dirac materials
may provide a microscopic mechanism to realize chiral Majorana hinge modes without
utilizing the proximity effect or external fields.
\end{abstract}
\maketitle

\paragraph{Introduction}

Majorana modes are the quasiparticles around a topological superconductor, and may
have the potential application in topological quantum computations \citep{Nayak2008Non,KiteavFault2003,Wilczek2009majorana,Alicea2012New,Beenakker2013Search,Elliott2015colloquium}.
Over the last two decades intensive efforts have been made to realize topological
superconductors \citep{Mourik2012signatures,Das2012Zero,Rokhinson2012fractional,Perge2014observation,Zhang2018Quantized,Sun2016Majorana,Wang2018Evidence,Lutchyn2018Evidence}.
The Majorana edge modes in a $p_{x}+ip_{y}$ spinless superconductor, a superconducting
analog of quantum Hall effect state, move in a dissipationless and unidirectional
way, i.e., are chiral, because of violation of time-reversal symmetry \citep{Ivanov2001non,Kitaev2001Unpaired,Volovik1999Fermion,Read2000Paired}.
As the $p$-wave superconductor is rare in nature, a hybrid system of quantum anomalous
Hall insulator and superconductor was alternatively proposed to realize the chiral
topological superconductor \citep{Qi2010chiral,Chung2011conductance,Wang2015chiral}.
However the existence of chiral Majorana modes are still inconclusive \citep{He2017science,Ji2018conductance,Huang2018disorder,Lian2018quantum,Kayyalha2020absence},
although several schemes for detection and application are proposed \citep{Strubi2011interference,Lian2018Topological,Li2019Majorana}.
Those proposals often rely heavily on the proximity effect or need an external magnetic
field to break the time reversal symmetry, which all make them difficult to be realized
in experiments. Very recently, a significant advance in the research of topological
quantum phases is a generalization to higher order topological insulators and superconductors
that can host localized modes near the corner, hinge or vertex of a system \citep{Benalcazar2017quantized,Benalcazar2017electric,Schindler2018higher,Zhu2019second,Langbehn2017reflection,Ezawa2018higher,Khalaf2018higher,Franca2018an,Calugaru2019higher,LiC20prb,Schindler2018higher2,Imhof2018topolectrical,Garcia2018observation,Peterson2018aquantized,Zhang2019second,Niu2020simulation}.
Several theoretical proposals have been put forward to realize the Majorana corner
modes in second-order topological superconductors \citep{Yan2018Majoran,Wang2018Weak,Wang2018HighTemperature,Zhu2019Higher,Hsu2018Majorana,Liu2018Majorana,Volpez2019Second,Zhang2019Helical,Zeng2019Majorana,Yan2019Higher,Zhang2020Detection,Pan2019Lattice,Franca2019Phase,Zhang2019Higher,Ahn2020Higher,Kheirkhah2020Majorana,Ghorashi2020Vortex,Hsu2020Inversion,Roy2020Higher,ZhangSB20prb}.
\textcolor{black}{In three dimensions, chiral Dirac modes can emerge along a hinge
between two surface planes on which the two gapped surface modes encounter when the
time-reversal symmetry is broken} \citep{Jack2019Observation,Peng2019Proximity,Yue2019Symmetry,Wu2020Inplane}.
This opens a new avenue to search chiral Majorana modes in topological materials.

Here we investigate all possible superconducting pairing channels in three-dimensional
(3D) massive Dirac materials with the $D_{2h}$ point group symmetry at the mean-field
level with long-ranged interactions. We find that the leading pairing channel can
be $s$-, $d$- or $s+id$-wave pairing by varying the relative strength of the intra-
and inter-orbital interactions. The $s$-wave pairing is topologically nontrivial
under time-reversal invariance and possesses gapless Majorana surface mode as proposed
by Fu and Berg \citep{Fu2010odd} and Sato \citep{Sato2010Topological}. For the
$s+id$-wave pairing channel, inclusion of $d_{xy}$- wave pairing breaks the time-reversal
symmetry and inversion symmetry, but preserves the combination of these two symmetries.
Consequently, the system becomes a second-order topological superconductor with chiral
Majorana hinge modes circulating along the four hinges parallel to the $z$-axis.
The topology behind chiral Majorana hinge modes can be characterized by a winding
number of the quadrupole moment, or the quantized quadrupole moment at the particle-hole
invariant momentum. This establishes a robust and new bulk-boundary correspondence
for the topological states of matter.

\paragraph{Model}

We investigate possible pairing channels in 3D Dirac materials with $D_{2h}$ point
group symmetry and time reversal symmetry by utilizing the symmetry analysis in the
irreducible representations of the group. The normal state Hamiltonian is given by
\begin{equation}
H_{0}=\sum_{\boldsymbol{k}}\Psi_{\boldsymbol{k}}^{\dagger}(h_{\boldsymbol{k}}-\mu)\tau_{z}\Psi_{\boldsymbol{k}}
\end{equation}
in the Nambu spinor basis $\Psi_{\boldsymbol{k}}=(\psi_{\boldsymbol{k}},\bar{\psi}_{\boldsymbol{k}})$,
where $\psi_{\boldsymbol{k}}$ is a four-component Dirac spinor, and $\bar{\psi}_{\boldsymbol{k}}=(-is_{y})(\psi_{-\boldsymbol{k}}^{\dagger})^{T}$
is its time-reversal hole partner. In the $\mathbf{k}\cdot\mathbf{p}$ theory, $h_{\boldsymbol{k}}=\sum_{i=x,y,z}v_{i}k_{i}\sigma_{x}s_{i}+m\sigma_{z}s_{0}$
with $v_{x,y,z}$ being the velocities along three directions, $m$ the Dirac mass,
and $\boldsymbol{s}$, $\boldsymbol{\sigma}$ and $\boldsymbol{\tau}$ the Pauli
matrices acting on spin, orbital, and Nambu space, respectively \citep{Shen2012Book,Shen2011Spin}
(We set $\hbar=1$.). Here $\mu$ is the chemical potential which is assumed to be
located in the conduction band. Furthermore, we consider the intra-orbital ($V_{0}=V_{z}=V_{intra}$)
and inter-orbital ($V_{x}=V_{y}=V_{inter}$) long-ranged attractive (density-density)
interaction between the Dirac fermions. By utilizing the Fierz identity \citep{Vafek2010prb,Savary2017prb,2018Venderbosprb,Note-on-SM},
the density-density product of four-fermion interaction can be decomposed into the
pairing terms
\begin{equation}
H_{int}=\sum_{\boldsymbol{k},\boldsymbol{k}^{\prime},i,j}\frac{V_{i}(\boldsymbol{k}-\boldsymbol{k}^{\prime})}{8\Omega}\left[\Psi_{\boldsymbol{k}}^{\dagger}\tau_{+}M_{ij}\Psi_{\boldsymbol{k}}\right]\left[\Psi_{\boldsymbol{k}^{\prime}}^{\dagger}\tau_{-}M_{ij}\Psi_{\boldsymbol{k}^{\prime}}\right]
\end{equation}
where $\Omega$ is the volume of the sample, $M_{ij}\equiv\sigma_{i}s_{j}$ and $\tau_{\pm}=\frac{1}{2}(\tau_{x}\pm i\tau_{y})$.
The interaction potential $V_{i}(\boldsymbol{k}-\boldsymbol{k}^{\prime})$ can be
decomposed by the Fermi-surface harmonics $\varphi_{\varGamma}(\boldsymbol{k})$
for each irreducible representation of the crystal point group $V_{i}(\boldsymbol{k}-\boldsymbol{k}^{\prime})=\sum_{\varGamma}V_{i}^{\varGamma}\varphi_{\varGamma}(\boldsymbol{k})[\varphi_{\varGamma}(\boldsymbol{k}^{\prime})]^{*}$
\citep{Nomoto2016Classification,Note-on-SM} where the sum over $\varGamma$ contains
all non-equivalent irreducible representations (IRs) of $D_{2h}$ group \citep{Dresselhaus2002Application}.
The basis functions for each IR are clearly not unique, and have been truncated to
the lowest order in momenta for simplicity. The calculation of the characteristic
coefficients for interaction $V_{i}^{\varGamma}$ can be can be found in Ref. \citep{Note-on-SM}.
The overall pairing functions are constructed by the orbital angular momentum part
$\varphi_{\varGamma}(\boldsymbol{k})$ and spin-orbital part $M_{ij}$ listed in
Table I. For the sake of simplicity, we only focus on the regime with the even $s$-
and $d$-wave pairing function {[}$\varphi_{\varGamma}(\boldsymbol{k})=\varphi_{\varGamma}(-\boldsymbol{k})${]}
for the orbital part, which restricts ourselves to the remaining six antisymmetric
pairing matrices $M_{ij}$, i.e., $[M_{ij}(-is_{y})]^{T}=-M_{ij}(-is_{y})$, due
to the Fermi-Dirac statistics. As a result of products of representations, the overall
paring function $\varphi_{\varGamma}(\boldsymbol{k})M_{ij}$ and the orbital angular
momentum part $\varphi_{\varGamma}(\boldsymbol{k})$ may belong to different IRs.

In a weak-coupling regime, the Cooper pairs emerge mainly near the Fermi surface.
Assuming that the Fermi level is located at the conduction bands, we can project
out the conduction bands, $\phi_{\boldsymbol{k}}=U_{\boldsymbol{k}}^{\dagger}\psi_{\boldsymbol{k}}$,
where $U_{\boldsymbol{k}}$ is a $4\times2$ matrix of conduction band eigenvectors
such that the projection operator for the conduction band $\mathcal{P}_{c}(\boldsymbol{k})=U_{\boldsymbol{k}}U_{\boldsymbol{k}}^{\dagger}=\frac{1}{2}(1+\frac{h_{\boldsymbol{k}}}{\epsilon_{\boldsymbol{k}}})$.
In this way, $H_{0}$ is diagonalized and the pairing functions are transformed to
$N_{\boldsymbol{k},\varGamma}=\varphi_{\varGamma^{\prime}}(\boldsymbol{k})U_{\boldsymbol{k}}^{\dagger}M_{ij}U_{\boldsymbol{k}}$
(see the fourth column in Table I). The superconducting pairing can be studied within
the framework of the functional integral method \citep{Altland2010condensed,Note-on-SM},
and the corresponding partition function takes the form $\mathcal{Z}=\int D[\varPhi,\varPhi^{\dagger}]e^{-\int_{0}^{\beta}d\tau\mathcal{L}[\varPhi,\varPhi^{\dagger}]}$
with the Lagrangian as $\mathcal{L}[\varPhi,\varPhi^{\dagger}]=\frac{1}{2}\sum_{\boldsymbol{k}}\varPhi_{\boldsymbol{k},\tau}^{\dagger}\tau_{0}\partial_{\tau}\varPhi_{\boldsymbol{k},\tau}+\widetilde{H}_{0}(\tau)+\widetilde{H}_{int}(\tau)-\mu\widetilde{N}(\tau)$,
where $\varPhi_{\boldsymbol{k}}=\left(\begin{array}{cc}
\phi_{\boldsymbol{k}}^{\dagger} & \bar{\phi}_{\boldsymbol{k}}^{\dagger}\end{array}\right)$ and $\widetilde{H}_{0}$, $\widetilde{H}_{int}$ and $\widetilde{N}$ are the projected
Hamiltonian, the interaction term and the particle density operator respectively.
The quartic interaction terms can be decoupled by performing the Hubbard-Stratonovich
transformation meanwhile the superconductivity order parameters $\eta_{\varGamma}$
are introduced. And the gap equations can be obtained by a variation of the action
with respect to the order parameters.

\begin{figure}
\includegraphics[width=8cm]{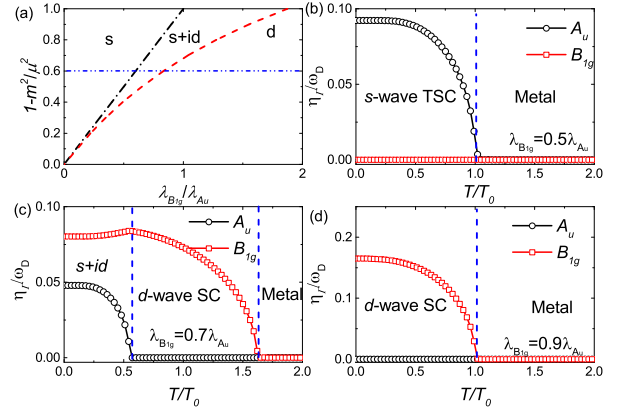}\caption{(a) Zero temperature phase diagram as a function of $1-m^{2}/\mu^{2}$ and $\lambda_{B_{1g}}/\lambda_{A_{u}}$.
The temperature dependence order parameter $\eta_{A_{u}}$ (b) for $s$-wave pairing
(black lines with open circles), $\eta_{B_{1g}}$ (d) for $d$-wave pairing (red
lines with open squares), and (c) for $s+id$-wave pairing. $\lambda_{A_{u}}$ is
set to be 1/2 and $1-m^{2}/\mu^{2}=0.6$. For regimes (b) and (d) with single order
parameter, the temperature is in the unit of its transition temperature $T_{0}=T_{c}^{\varGamma}$.
For the mixed pairing regime (c), $T_{0}=T_{c}^{A_{u}}$.\label{fig:Zero_temperature_phase_diagram}}
\end{figure}

\begin{table}
\begin{tabular}{ccccc}
\toprule 
IRs & orbital & $\varphi_{\varGamma}(\boldsymbol{k})M_{ij}$ & $N_{\boldsymbol{k},\varGamma}$ & $\langle N_{\boldsymbol{k},\varGamma}^{2}\rangle$\tabularnewline
\midrule
\midrule 
$A_{g}$ & intra & $\varphi_{0}\sigma_{0}s_{0}$ & $\varphi_{0}\widetilde{\sigma}_{0}$ & $1$\tabularnewline
\midrule 
$B_{1g}$ & intra & $\varphi_{xy}\sigma_{0}s_{0}$ & $\varphi_{xy}\widetilde{\sigma}_{0}$ & $1$\tabularnewline
\midrule 
$B_{2g}$ & intra & $\varphi_{xz}\sigma_{0}s_{0}$ & $\varphi_{xz}\widetilde{\sigma}_{0}$ & $1$\tabularnewline
\midrule 
$B_{3g}$ & intra & $\varphi_{yz}\sigma_{0}s_{0}$ & $\varphi_{yz}\widetilde{\sigma}_{0}$ & $1$\tabularnewline
\midrule 
$A_{u}$ & inter & $\varphi_{0}\sigma_{x}s_{0}$ & $\varphi_{0}\frac{\boldsymbol{p}\cdot\widetilde{\boldsymbol{\sigma}}}{\epsilon_{\boldsymbol{k}}}$ & $1-\frac{m^{2}}{\mu^{2}}$\tabularnewline
\midrule 
$B_{1u}$ & inter & $\varphi_{0}\sigma_{y}s_{z}$ & $\varphi_{0}\frac{(\boldsymbol{p}\times\widetilde{\boldsymbol{\sigma}})_{z}}{\epsilon_{\boldsymbol{k}}}$ & $\frac{2}{3}(1-\frac{m^{2}}{\mu^{2}})$\tabularnewline
\midrule 
$B_{2u}$ & inter & $\varphi_{0}\sigma_{y}s_{y}$ & $\varphi_{0}\frac{(\boldsymbol{p}\times\widetilde{\boldsymbol{\sigma}})_{y}}{\epsilon_{\boldsymbol{k}}}$ & $\frac{2}{3}(1-\frac{m^{2}}{\mu^{2}})$\tabularnewline
\midrule 
$B_{3u}$ & inter & $\varphi_{0}\sigma_{y}s_{x}$ & $\varphi_{0}\frac{(\boldsymbol{p}\times\widetilde{\boldsymbol{\sigma}})_{x}}{\epsilon_{\boldsymbol{k}}}$ & $\frac{2}{3}(1-\frac{m^{2}}{\mu^{2}})$\tabularnewline
\bottomrule
\end{tabular}

\caption{Eight classes of the basis of the pairing functions according to $D_{2h}$ point
group symmetry. From left to right, each column shows the irreducible representations
(IRs) of the point group $D_{2h}$, intra-orbital or inter-orbital pairing, the pairing
channels $\varphi_{\varGamma}(\boldsymbol{k})M_{ij}$ with antisymmetric matrices,
the pairing functions $N_{\boldsymbol{k},\varGamma}$ by projecting onto the states
close to the Fermi surface, and the average over the Fermi surface $\langle N_{\boldsymbol{k},\varGamma}^{2}\rangle$
with $\langle..\rangle$ stands for the Fermi surface average for arbitrary function
$\langle...\rangle=\sum_{\boldsymbol{k}}...\delta(\epsilon_{\boldsymbol{k}}-\mu)/\sum_{\boldsymbol{k}}\delta(\epsilon_{\boldsymbol{k}}-\mu)$
with $\epsilon_{\boldsymbol{k}}=\sqrt{\sum_{i}v_{i}^{2}k_{i}^{2}+m^{2}}$. The Pauli
matrices $\widetilde{\sigma}_{i}$ denote the Kramers-degenerated conduction bands
and we have introduced the notation $\boldsymbol{p}=(v_{x}k_{x},v_{y}k_{y},v_{z}k_{z})$
. The Fermi-surface harmonics $\varphi_{0}=1$ and $\varphi_{ij}=\frac{\sqrt{15}}{1-m^{2}/\mu^{2}}\frac{v_{i}v_{j}k_{i}k_{j}}{\epsilon_{\boldsymbol{k}}^{2}}$
(for $i,j=x,y,z$ and $i\protect\ne j$).}
\end{table}

\begin{figure}
\includegraphics[width=8cm]{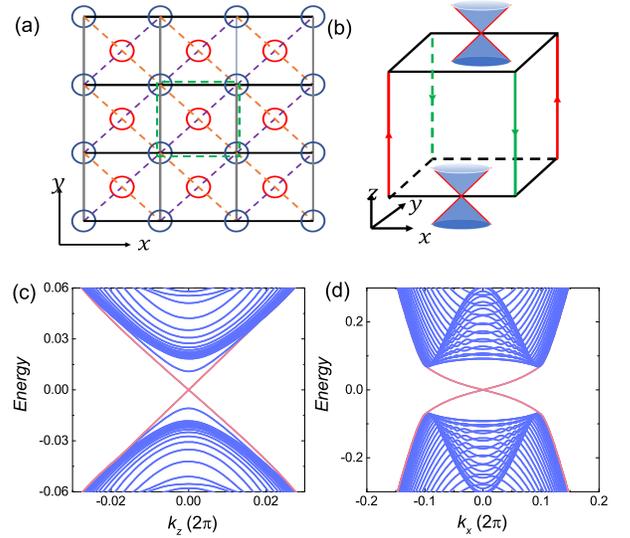}\caption{(a) Schematic of cross section lattice and the boundary of the x-y plane. The unit
cell (the dash green box) consists of two sublattices indicated by blue and red circles.
(b) The schematic of Majorana hinge modes and gapless Majorana surface modes in the
case of $s+id$-wave pairing. (c) The dispersion spectrum of Majorana hinge modes
for quasi-1D hinges along $z$ direction with $L_{x}=L_{y}=60$. (d) The dispersion
spectrum of the Majorana surface modes in the x-y plane. The open boundary condition
is adopted along z direction with the height $L_{z}=200$ and the periodic boundaries
are adopted along $x$ and $y$ directions(see Sec. VII of Ref. \citep{Note-on-SM}).
The chemical potential $\mu=0.5$. \label{fig:latticeband}}
\end{figure}

\paragraph{Determination of the pairing symmetry}

Now, we evaluate the transition temperature $T_{c}^{\varGamma}$ for each pairing
channel listed in Table I. It can be obtained by solving the linearized gap equation
near the transition temperature $T_{c}$ at which the order parameter is vanishingly
small. Generally speaking, $T_{c}^{\varGamma}\simeq1.13\omega_{D}\exp\left[-\frac{1}{\lambda_{\varGamma}\langle N_{\boldsymbol{k},\varGamma}^{2}\rangle}\right]$
associated with each IR can be different. The transition temperature is dictated
by two factors. The first one is the average of the square of the projected paring
matrices $\langle N_{\boldsymbol{k},\varGamma}^{2}\rangle$ over the Fermi surface.
The second one is the dimensionless coupling strength $\lambda_{\varGamma}=2g_{\varGamma}\rho(\mu)$
with the effective interaction $g_{\varGamma}$ near the Fermi level and the density
of states for the normal state $\rho(\mu)$. The inter-orbital interaction $V_{i=x,y}(\boldsymbol{r}-\boldsymbol{r}^{\prime})$
and the intra-orbital interaction $V_{i=0,z}(\boldsymbol{r}-\boldsymbol{r}^{\prime})$
give rise to pairing in the odd-parity channels $(A_{u},B_{1u},B_{2u},B_{3u})$ and
even-parity channels ($A_{g},B_{1g},B_{2g},B_{3g}$), respectively. The four odd-parity
pairing channels are generated from the local inter-orbital interaction and the transition
temperatures satisfy $T_{c}^{A_{u}}\gg T_{c}^{B_{1u}},T_{c}^{B_{2u}},T_{c}^{B_{3u}}$.
Thus we need only to consider the $s$-wave superconductivity belonging to the $A_{u}$
representation among the four odd-parity pairing channels. In evaluating $d$-wave
pairing matrices $\langle N_{\boldsymbol{k},B_{1g}}^{2}\rangle$, $\langle N_{\boldsymbol{k},B_{2g}}^{2}\rangle$,
and $\langle N_{\boldsymbol{k},B_{3g}}^{2}\rangle$, it is convenient to perform
rescaling coordinate transformation $\widetilde{r}_{i}=\zeta_{i}r_{i}$ for $i=x,y,z$
with the anisotropic factor $\zeta_{i}=(v_{x}v_{y}v_{z})^{1/3}/v_{i}$, and we use
$\zeta_{x}=1$, $\zeta_{y}=2$ and $\zeta_{z}=1/(\zeta_{x}\zeta_{y})=1/2$ throughout
this work. Under the rescaling procedure, the Fermi surface average of the $d-$wave
paring matrices can be obtained analytically as $\langle N_{\boldsymbol{k},B_{1g}}^{2}\rangle=\langle N_{\boldsymbol{k},B_{2g}}^{2}\rangle=\langle N_{\boldsymbol{k},B_{3g}}^{2}\rangle=1$.
Without loss of generality, we assume $g_{B_{1g}}>g_{B_{2g}}>g_{B_{3g}}$. Under
this condition, it is unlikely for the system to form the order parameter in $B_{2g}$
and $B_{3g}$ channels. Depending on the pairing interaction, we cannot avoid the
possibility of the pairing belonging to the $A_{g}$ representation. Since this order
parameter breaks no additional symmetries besides the $U(1)$ gauge symmetry, we
disregard this conventional pairing for the further discussions. We have also checked
numerically that the conclusion remains unchanged even with the inclusion of this
pairing.

From now on, we focus on the topological nontrivial phases with $s$- and $d$-wave
pairing which belong to two different irreducible representations $A_{u}$ and $B_{1g}$,
respectively. Generally, $\eta_{\varGamma}$ are complex which can be parameterized
as $\eta_{\varGamma}=|\eta_{\varGamma}|e^{i\phi_{\varGamma}}$ where $|\eta_{\varGamma}|$
and $\phi_{\varGamma}$ are real. We consider the case that the phase difference
for $s$-wave and $d$- wave pairing $\Delta\phi=\phi_{A_{u}}-\phi_{B_{1g}}$, and
the order parameter for $s$-wave pairing is real, i.e. $\phi_{A_{u}}=0$. By minimizing
the free energy with the two superconducting order parameters, the relative phase
should to be $\Delta\phi=\pm\pi/2$ if they coexist \citep{Note-on-SM}. Then the
pairing function can be expressed as $\sum_{\varGamma}\eta_{\varGamma}N_{\boldsymbol{k}}^{\varGamma}=\eta_{A_{u}}N_{\boldsymbol{k},A_{u}}-i\eta_{B_{1g}}N_{\boldsymbol{k},B_{1g}}$,
with $N_{\boldsymbol{k},A_{u}}=\varphi_{0}\frac{\boldsymbol{p}\cdot\widetilde{\boldsymbol{\sigma}}}{\epsilon_{\boldsymbol{k}}}$
and $N_{\boldsymbol{k},B_{1g}}=\varphi_{xy}\widetilde{\sigma}_{0}$. Thus, the projected
Bogoliubov-de Gennes (BdG) Hamiltonian can be expressed as,
\begin{equation}
\widetilde{h}_{\boldsymbol{k}}^{BdG}=(\epsilon_{\boldsymbol{k}}-\mu)\widetilde{\sigma}_{0}\tau_{z}+\frac{\eta_{A_{u}}}{\mu}\boldsymbol{p}\cdot\widetilde{\boldsymbol{\sigma}}\tau_{x}+\frac{\sqrt{15}\eta_{B_{1g}}v_{x}v_{y}k_{x}k_{y}}{\mu^{2}-m^{2}}\widetilde{\sigma}_{0}\tau_{y}.\label{eq:projectedBdG}
\end{equation}
The gap equations are reduced to a pair of coupled self-consistent equations of superconducting
gap for the two paring amplitudes $\eta_{A_{u}}(T)$ and $\eta_{B_{1g}}(T)$.

\paragraph{$s+id$ wave pairing state}

Now we turn to explore the possibility of a mixed $s+id$-wave pairing which breaks
the time-reversal symmetry spontaneously. Figure \ref{fig:Zero_temperature_phase_diagram}(a)
shows the phase diagram at zero temperature as function of $1-m^{2}/\mu^{2}$ and
the interaction ratio $\lambda_{B_{1g}}/\lambda_{A_{u}}$ with fixed $\lambda_{A_{u}}$
to illustrate the the competition between the $s$- and $d$-wave pairing superconductivity.
The red and black dotted lines indicate the phase boundary separating the purely
$s$ ($\eta_{A_{u}}(0)\ne0,\eta_{B_{1g}}(0)=0$) or $d$-wave pairing $(\eta_{A_{u}}(0)=0,\eta_{B_{1g}}(0)\ne0)$
and the mixed $s+id$-wave pairing $(\eta_{A_{u}}(0),\eta_{B_{1g}}(0)\ne0)$. The
competition between the $s$- and $d$-wave pairing channels can lead to either purely
$s$- or $d$-wave pairing state, or a mixed $s+id$-wave state. It is intuitively
clear that for such an $s+id$-solution to be held, pairing strengths $\lambda_{A_{u}}$
and $\lambda_{B_{1g}}$ need to be comparable: otherwise, a s-wave or a d-wave will
dominate. Adjusting the chemical potential toward the band edge, the region for the
mixed pairing shrinks. Thus, the mixed pairing is more likely to occur in a system
when the chemical potential locates away from the band edge.

Then we discuss the behaviors of two different order parameters $\eta_{A_{u}}$ and
$\eta_{B_{1g}}$ at finite temperature, which are calculated by solving the gap equations
as a function of temperature for several values of $\lambda_{B_{1g}}/\lambda_{A_{u}}$
and fixed $1-m^{2}/\mu^{2}=0.6$. As shown in Fig. \ref{fig:Zero_temperature_phase_diagram}(b)(d),
if $\lambda_{B_{1g}}/\lambda_{A_{u}}<r_{max}=[(1-m^{2}/\mu^{2})^{-1}-\frac{14}{15}\lambda_{A_{u}}]^{-1}$
or $\lambda_{B_{1g}}/\lambda_{A_{u}}>r_{min}=1-m^{2}/\mu^{2}$ \citep{Note-on-SM},
it is a pure $s$ or $d$-wave superconductivity whose critical temperature is precisely
determined by $T_{c}^{\varGamma}$, that the superconducting transition is only specific
to one of the IRs. The mixed paired state appears at the intermediate region $\lambda_{B_{1g}}/\lambda_{A_{u}}\in(r_{min},r_{max})$.
As shown in Fig. 1(c), for $\lambda_{B_{1g}}/\lambda_{A_{u}}=0.7$ ( $\in(r_{min},r_{max})$),
as the temperature decreases down to a certain value $\sim1.63T_{0}$, $\eta_{B_{1g}}$
first appears. After then, the $s$ wave pairing $\eta_{A_{u}}$ appears and $\eta_{B_{1g}}$
increases gradually with temperature until $\sim0.55T_{0}$. After that, the $d$-wave
component reduces while the $s$-wave component grows up as temperature decreases
to zero. This indicates that, with decreasing the temperature, it undergoes a topological
phase transition from pure $d$-wave superconductivity to $s+id$-wave superconductivity
in specific conditions.

\paragraph{Majorana hinge and surface modes in the $s+id$ wave pairing state}

The $s+id$-wave pairing superconducting state is higher-order topologically nontrivial,
which is revealed from the existence of Majorana hinge and surface modes. We adopt
the tight-binding approximation on a cubic lattice with the lattice orientation of
the x-y plane as shown in Fig. \ref{fig:latticeband}(a)\citep{Note-on-SM}. Four
chiral Majorana hinge modes and two Majorana surface modes are illustrated in Fig.
\ref{fig:latticeband}(b). The origin of the topological hinge modes in $s+id$-wave
pairing state can be heuristically explained in the following picture. According
to the odd-parity superconductivity criterion \citep{Fu2010odd}, the system of the
single $s$-wave paring ($A_{u}$ ) should be a time-reversal invariant topological
superconductor with massless helical Majorana modes on the surface. The inclusion
of the $d$-wave order parameter gaps out the Dirac cones of the Majorana surface
modes on the surfaces parallel to the z-axis as its relative $\pi/2$ phase to the
$s$-wave order parameter breaks the time reversal symmetry. However, the $d$-wave
pairing gap function vanishes at the mirror planes and acts as mass domain walls
for Majorana surface modes. Consequently, the chiral Majorana hinge modes are formed
around the domain wall or along the hinges. The dispersion spectra for the Majorana
hinge states as a function of $k_{z}$ are presented in Fig. \ref{fig:latticeband}(c).
The gapless Majorana hinge modes in the gap are marked by the red lines and localized
near the four hinges along the z-direction. Their dispersions are linear in $k_{z}$,
$E_{hinge}=\pm\frac{\eta_{A_{u}}}{\mu}v_{z}k_{z}$, obtained from the projected BdG
Hamiltonian in Eq. (\ref{eq:projectedBdG}). On the top and bottom surface of the
x-y plane, the $d$-wave pairing breaks the time-reversal symmetry, but does not
open the band gap of surface Majorana states. The dispersion spectra are plotted
in Fig. \ref{fig:latticeband}(d), in which the gapless Majorana surface states are
marked by the red lines. The BdG Hamiltonian with the odd-parity $s$-wave pairing
$(\propto\sigma_{x}\tau_{x})$ possesses the time-reversal symmetry $\mathcal{T}=is_{y}\mathcal{K}$
and the inversion symmetry $\widetilde{\mathcal{I}}\equiv\mathcal{I}\tau_{z}=\sigma_{z}\tau_{z}$.
The inclusion of the $d$-wave pairing $(\propto k_{x}k_{y}\tau_{y})$ breaks either
$\mathcal{T}$or $\widetilde{\mathcal{I}}$, but preserves $\widetilde{\mathcal{I}}\mathcal{T}$.
This combined symmetry makes the Majorana hinge modes at diagonal hinges and the
Majorana surface states at the opposite surface related by the symmetry operation
$\widetilde{\mathcal{I}}\mathcal{T}$.

\begin{figure}
\includegraphics[width=8cm]{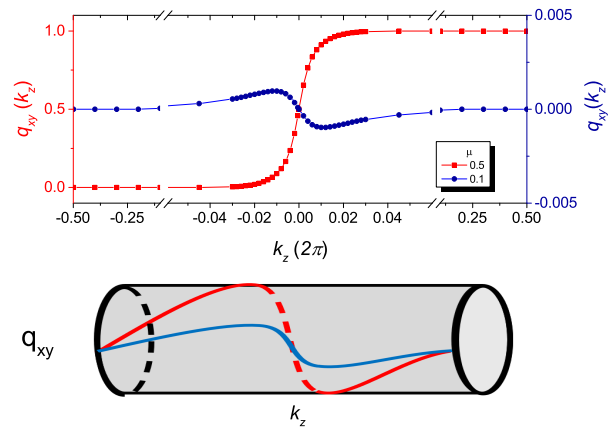}

\caption{The quadrupole moment $q_{xy}(k_{z})$ as a function of $k_{z}$ for $\mu=0.5$ (red)
with the winding number $\Delta q_{xy}=1$ and $0.1$ (blue) with $\Delta q_{xy}=0$.
The bottom: $q_{xy}$ is rolled as a tube with $q_{xy}$ module 1. The other parameters
used are the same as Fig. \ref{fig:latticeband}. \label{fig:The-quadrupole-moment}}
\end{figure}

\paragraph{Topological invariants}

Now we come to discuss the topological invariant which is related to the Majorana
hinge modes. We take the periodic condition along the z axis such that $k_{z}$ is
a good quantum number. The quadrupole moment for each $k_{z}$ is given by \citep{Li2020,Kang2019Many,Wheeler2019Many}
\begin{equation}
q_{xy}(k_{z})=\frac{1}{2\pi}\mathrm{Im}\log\left[\mathrm{Det}[U_{k_{z}}^{\dagger}QU_{k_{z}}]\sqrt{\mathrm{Det}Q^{\dagger}}\right],
\end{equation}
where the matrix $U_{k_{z}}$ is constructed by the occupied ground states, $Q=e^{2\pi i\hat{x}\hat{y}/L_{x}L_{y}}$,
$\hat{x}$ and $\hat{y}$ are the position operators, and $L_{x}$ and $L_{y}$ the
length of our system in x and y direction, respectively. Generally the particle-hole
symmetry $\mathcal{P}$ is broken for a specific $k_{z}$ as the two states at $k_{z}$
and $-k_{z}$ are connected by the $\mathcal{P}$ symmetry. However the symmetry
restores at $k_{z}=0$ and $\pi$ (half of the reciprocal lattice vector), i.e.,
the particle-hole invariant momentum. At these momenta, we can prove that the quadrupole
moment $q_{xy}$ is quantized to be 0 or 1/2 module 1 \citep{Note-on-SM}. Quantized
quadrupole moment of $q_{xy}=1/2$ means the existence of the corner states of zero
energy, a signature of the second-order topological phase \citep{Benalcazar2017electric,Benalcazar2017quantized}.
The quantization is removed once $k_{z}$ moves away the invariant momentum. Additions
of the corner states for different $k_{z}$ evolve into the chiral hinge states in
three dimensions. Furthermore, the $\mathcal{P}$ symmetry connects the unoccupied
states at momentum $k_{z}$ with unoccupied states at $-k_{z}$, which gives $q_{xy}(k_{z})+q_{xy}(-k_{z})=0$
for a trivial case and 1 for a nontrivial case. Thus in the $s+id$-wave pairing
state, a winding number can be introduced $\Delta q_{xy}=\int_{-\pi}^{\pi}dk_{z}\partial_{k_{z}}q_{xy}(k_{z}).$
Two $k_{z}$-dependent quadrupole moments for trivial (blue line with circle) and
nontrivial (red line with square) cases are plotted in Fig. \ref{eq:projectedBdG}.
For $\mu=0.5$, it is found that $q_{xy}=1/2$ at $k_{z}=0$, and 0 at $k_{z}=\pi$.
The winding number $\Delta q_{xy}=1$, which means the state is topologically nontrivial
and is related to the presence of chiral Majorana hinge modes. In this way, we establish
a bulk-hinge correspondence in the topologically nontrivial superconducting state.
\begin{acknowledgments}
This work was supported by the Ministry of Science and Technology of China under
Grant No. 2019YFA0308603, Natural Science Foundation of China under Grant No. 11774317,
and Natural Science Foundation of Zhejiang under Grant No. LQ20A040003.
\end{acknowledgments}

\end{document}